\documentclass[aps,prl,twocolumn,10pt,showpacs,floatfix]{revtex4-1}
\usepackage{amsmath}
\usepackage{amssymb}
\usepackage{amsthm}
\usepackage{latexsym}
\usepackage{hyperref}
\usepackage{cleveref}
\usepackage{graphicx}
\hypersetup{
    colorlinks,
    citecolor=blue,
    linkcolor=blue,     urlcolor=blue
}

\begin{document}
 
\title{X-ray amplification from a Raman Free Electron Laser}
\author{I.A. Andriyash}
\email{andriyash@celia.u-bordeaux1.fr}
\author{E. d'Humi\`eres}
\author{V.T. Tikhonchuk}
\author{Ph. Balcou}
\affiliation{Univ. Bordeaux, CNRS,  CEA, CELIA (Centre Lasers Intenses et Applications), UMR 5107,  F33400 Talence, France}
 
\begin{abstract}
We demonstrate that a mm-scale free electron laser can operate in the X-ray range, in the interaction between a
moderately relativistic electron bunch, and a transverse high intensity optical lattice. The corrugated light-induced
ponderomotive potential acts simultaneously as a guide and as a low-frequency wiggler, triggering stimulated Raman
scattering. The gain law in the small signal regime is derived  in a fluid approach, and confirmed from Particle-In-Cell
simulations. We describe the nature of bunching, and discuss the saturation properties. The resulting all-optical Raman
X-ray laser opens perspectives for ultra-compact coherent light sources up to the hard X-ray range. \end{abstract}

\maketitle

The advent of fully coherent light  sources in the X-ray range promises to be the next revolution in X-ray science,
leading to many scientific, industrial and health applications.  Large scale X-ray free electron laser (XFEL) projects
have been launched, and start supplying high brightness beams for novel physics experiments \cite{Emma:NaturPhoton}.
However, these large scale infrastructure cannot  allow for widespread dissemination of the  XFEL technologies,
which would require to shorten very significantly the length of the linear accelerators and magnetic undulators.
This size constraints also prevent  to reach the hard X-ray range, which would open entirely new perspectives.

Few alternative strategies have been proposed to supply compact X-ray free electron lasers, based on  substitution of
the LINAC by laser-wakefield acceleration \cite{gruener1,nakajima},  use of original compact undulators as an ion
channel \cite{Whittum:PRL1990}, or  a counter-propagating laser \cite{Gea-Banacloche:QE,Dobiasch:QE,petrillo}. Laser
undulators offer indeed a key advantage :  laser wavelengths at the $\mu$m level allow one to reach X-ray photon
energies with moderately relativistic electrons, of kinetic energies of few tens of MeV only. In all cases, the XFEL is
expected to operate in the conventional regime of stimulated inverse Compton scattering, imposing severe limitations on
electron emittance, transport, kinetic energy spread, and laser uniformity for laser undulators
\cite{sprangle:PhysRevSTAB}, thus hindering  prospects of experimental demonstrations.

\begin{figure}[!ht]\centering
\includegraphics[width=0.45\textwidth]{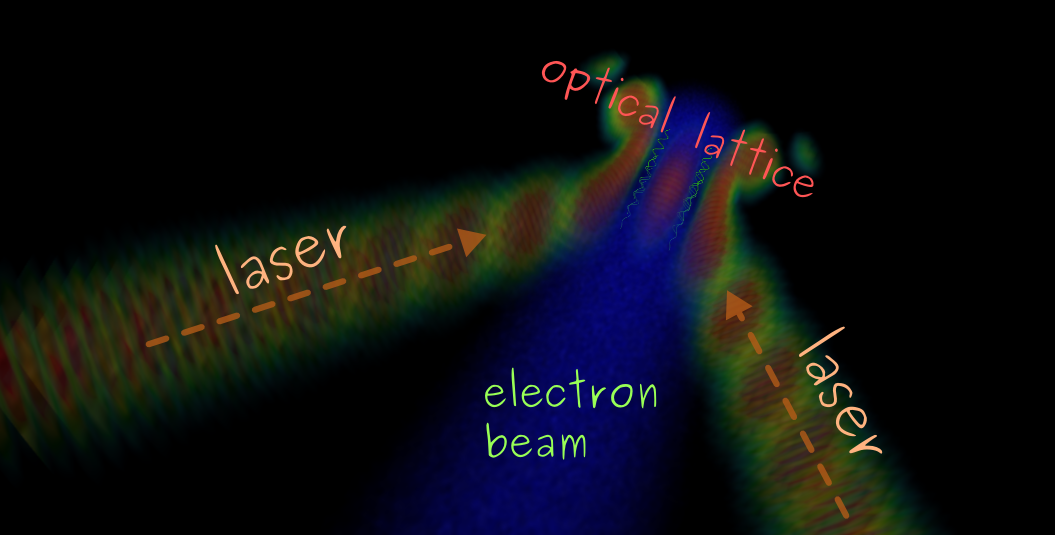}\\
\caption{X-ray Raman scattering geometry in a reference system moving with the electron bunch.} \label{fig1}
\end{figure}

A conceptually different new scheme considers a relativistic electron bunch injected into the overlap
region between two transversally incident, counter-propagating intense lasers beams
\cite{balcou:epjd}. The setup is depicted in Fig. 1, directly in the  reference frame of the electron bunch. The
interference between the  laser beams forms an optical lattice, and induces a spatially corrugated ponderomotive
potential for the incident electrons that is trapping them transversely. The electron dynamics then consists of
high frequency oscillations induced by the two lasers along the laser polarization direction, and of low frequency
oscillations along the interference direction, similar to betatron oscillations, with a characteristic angular 
frequency $\Omega$. Light is hence scattered at the betatron frequency, and on the Stokes and anti-Stokes lines around
the laser frequency. In the laboratory frame, this scattering is Doppler up-shifted by $2\gamma^2$, where $\gamma$ is
the electron Lorentz factor. Scattering is spontaneous as long as the electron motions are uncorrelated; however,
electrons may also exhibit a collective low-frequency oscillatory behaviour, so that we can expect a stimulated Raman
instability and coherent emission of X-ray radiation in the forward direction. This Raman-type scattering should be
distinguished from the known Raman instabilities in conventional long wavelength Free Electron Lasers, where the system
oscillations are the Langmuir plasma waves \cite{renz}. This new Raman instability dominates if the bounce frequency
$\Omega$ is greater than the electron beam plasma frequency $\omega_p$.

The aim of the present Letter is to investigate  the  X-ray amplification process in this trapped mode on the basis
of particle-in-cell numerical simulations, and to derive its main scaling laws on small-signal gain and saturation, from
an analytical hydrodynamic model.

Let us consider a relativistic electron bunch moving along a  $z$-axis, with an average Lorentz factor up to a hundred
typically, and number density $n_e^0$,  incident onto the optical lattice resulting from overlapping twin laser beams,
interfering  along the $x$-axis. The light lattice intensity is considered as  sub-relativistic, typically in the range
$10^{16}$ to $10^{18}$ W/cm$^2$ for near infrared lasers.

The process is  conveniently described in the  frame moving with the electron beam velocity ${v_b = c \beta_b \lesssim
1}$, for  which the incident laser and scattered field frequencies are similar, and electron motion can be treated
non-relativistically. The twin lasers appear in that frame at oblique incidence, with an angle defined as
\({k_{0\perp}/k_{0\parallel}= 1/\gamma \beta_b\ll 1}\), see \cref{fig1}. Sub-relativistic lattice intensity defines
a moderate vector potential, which is a Lorentz invariant and in dimensionless units reads $a_0 = 0.85  \times 10^{-9}
\lambda_0^l \sqrt{I^l}$, where $I$ (W/cm$^2$) is the laser intensity and $\lambda_0^l$ ($\mu$m) the laser wavelength.
With the superscript ``\textit{l}'' we denote the parameters in the laboratory reference frame.

In that frame, electrons first enter a ramp-up region, followed by an optical lattice of a constant intensity;
in the bunch frame, this process is described as a relatively slow switch-on of the external electromagnetic field. The
normalized vector potential of the incident  laser irradiation reads as
\begin{equation}\label{pump_func}
a_L= a_1 + a_2 = 2 a_0(t) \,\sin(k_{0\perp} x) \,\cos(\omega_0 t +k_{0\parallel}z)\,,
\end{equation}
resulting into a ponderomotive potential $U_p$: 
\[U_p(x,t)= m_e c^2 a_0(t)^2 \sin^2 (k_{0\perp}x) \,.\]
Near the bottom of the potential, electrons oscillate with the bounce frequency $\Omega(t) = \sqrt{2} a_0(t) k_{0\perp}
c$. 

Field amplitude grows during a time $t_{inj}$ up to $a_0$. Electrons with transverse velocities, $v_\perp\le \sqrt{2}a_0
c $, may be trapped in the ponderomotive potential and oscillate along the closed phase trajectories in $(x,p_x)$-plane.
Particles with lower initial velocities are trapped before $t_{inj}$, and further growth of potential adiabatically
compresses their phase trajectories, thus increasing the velocity amplitude and electron density at the beam axis
\cite{andriyashEPJD2011}. The maximum electron density after injection can be estimated as:
\begin{equation}
n_e= n_e^0 (\sqrt{2}a_0/\delta\beta_{0\perp})\,.
\end{equation}

Trapped electrons are driven at high frequency by the lasers, and oscillate transversely in the optical lattice
potential, akin to the betatron oscillations  in plasma ion channels \cite{Pukhov:PRL04,taphuocPOP05}, with a harmonic
law $x_e = x_m\cos(\Omega t)$, where $\Omega= \sqrt{2} a_0 k_{0\perp}c$. Radiation is emitted in two  frequency
ranges : at the betatron frequency \cite{Sepke:PRE026501,fedorov:353}; and symmetrically  around
the laser frequency,  at the Stokes $\omega_0 - \Omega$ and anti-Stokes $\omega_0 + \Omega$ frequencies
\cite{balcou:epjd}. 

\begin{figure}[!ht]\centering
\includegraphics[width=0.45\textwidth]{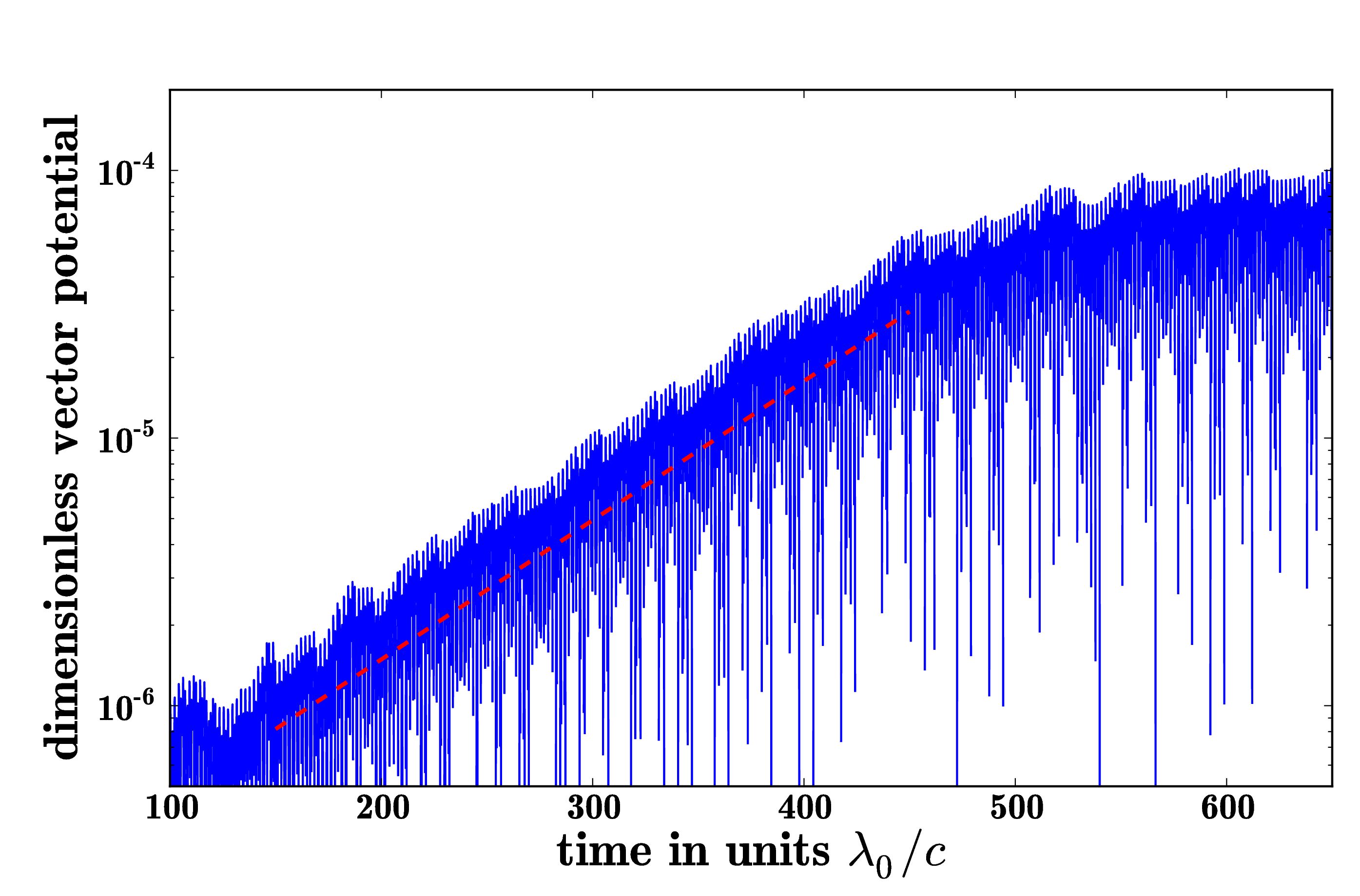}\
\caption{Evolution of vector potential at the center of right boundary $(z,x) = (l_z,0)$. The exponential fit
$\Gamma/\omega_0 = 1.8\cdot 10^{-3}$ is shown by a red dotted line.}\label{fig2}
\end{figure}

We have developed a  numerical tool, dubbed EWOK, to investigate scattering process in the bunch frame. The
two-dimensional code is based on a general particle-in-cell approach \cite{langdon04}; it factors out the
high-frequency, laser-induced oscillations of electrons, solving relativistic motion equations for macro-particles in
electrostatic and ponderomotive potentials. The scattered wave is described in the envelope approximation
${a_s=\bar{a}\exp(-i\omega_0 t+ik_0z)}$ as a slowly oscillating amplitude $\bar{a}$. This allows us to reduce to
first order the propagation equation along $z$, and to solve it over $x$ by a Fourier transform \({\partial_x^2
\bar{a}\to -k_x^2 \bar{a}_{k_x}}\). The simulation domain is a rectangular box, whose dimensions $l_{z,x}$ define a
discrete number of electromagnetic eigenmodes. Boundary conditions are periodic for both electromagnetic waves, while
for the particles, boundaries are periodic along z-axis and absorbing along x-direction. Thus, the beat between the
laser and the scattered wave must be periodic, which imposes constraints on the wave-vectors $(k_s-k_0)$ of the
scattered wave, and $(k_s+k_{0\parallel})$ of the longitudinal ponderomotive force. For a plane scattered wave at the
Stokes frequency, the corresponding periodicity conditions read:
\begin{align}\label{eig_conds} 
&k_s = k_0 (1-N_1\lambda_0/l_z)\\
&l_z/\lambda_0 = N_2/(1+\beta_b)\,.\nonumber
\end{align}
$N_1$ and $N_2$ being integers. These conditions lead to exceedingly long simulation domains for large $\gamma$ values.
We perform therefore simulations with a relatively low $\gamma$, and exploit the analytical laws to rescale the
results to experimentally relevant parameters. 

We have used EWOK to simulate amplification on the Stokes mode, satisfying the resonance condition ${\omega_0 - \omega_s
= \Omega}$, with physical parameters summarized in \cref{tab2}, and on the basis of
$5.2\cdot 10^6$ macro-particles.

\begin{figure}[!ht]\centering
 \includegraphics[width=0.45\textwidth]{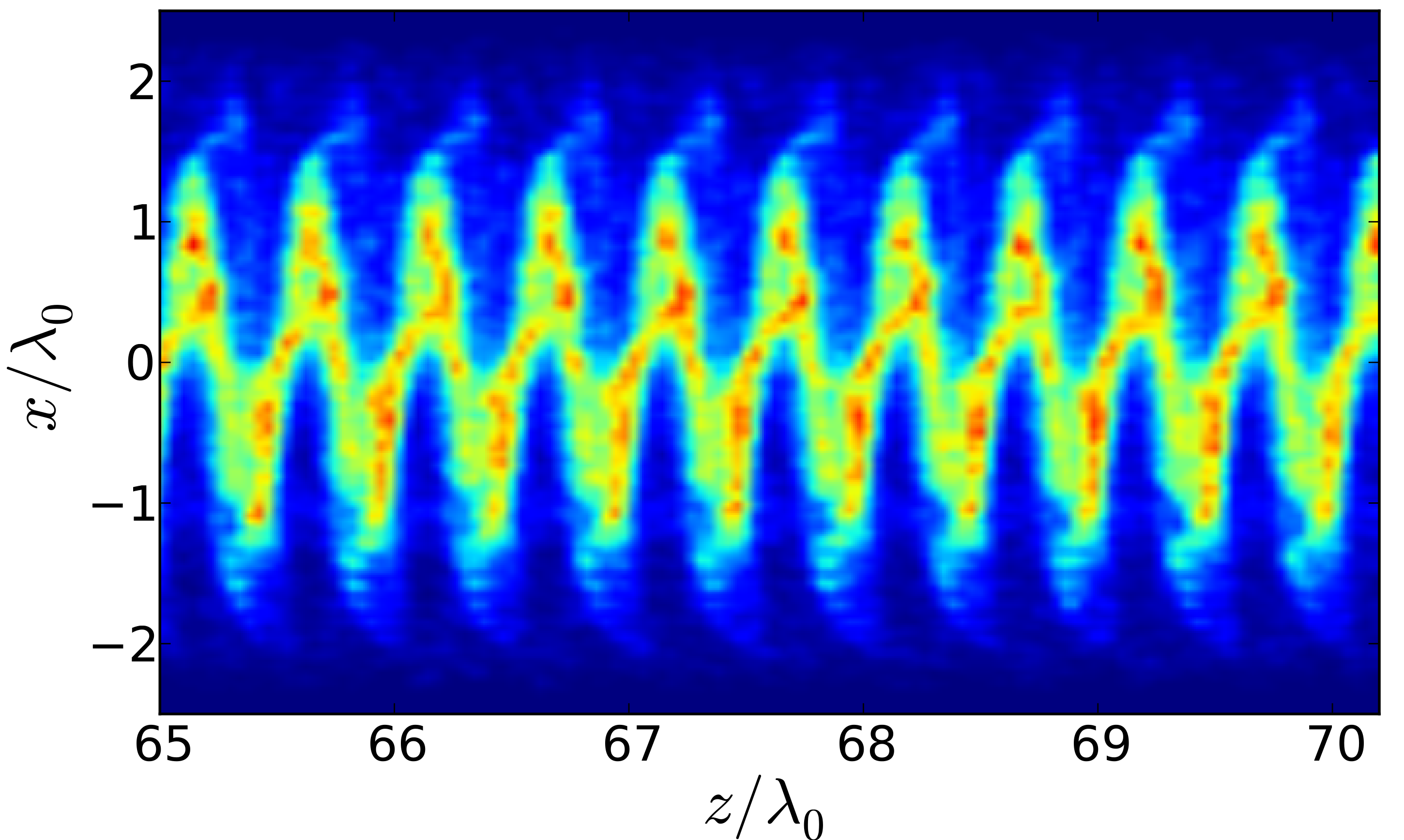}\\
 \caption{Snapshot of electron density distribution at saturation time $600\lambda_0/c$.}\label{fig3}
\end{figure}

\begin{table}[!ht]\centering
\caption{Interaction parameters}\label{tab2}
\begin{tabular*}{0.45\textwidth}{p{1.5cm} r  | p{1.5cm}  r}
\hline \hline
\multicolumn{2}{c}{laboratory } & \multicolumn{2}{c}{moving beam}\\
\hline
\multicolumn{4}{c}{electron beam} \\
\hline
current & $8$ kA & density & $2\cdot 10^{-5}\, n_c$ \\
width & $3\mu$m & width & $3\mu$m\\
duration & $10$ fs & length& $30\,\mu$m  \\
$\delta\gamma_e/\gamma$ & $10^{-4}$ & $\delta\beta_{\parallel0}$ & $10^{-4}$ \\
emittance & $0.3$ mm$\cdot$mrad & $\delta\beta_{\perp0}$ & $0.1$\\
energy & $4.7$ MeV & $\gamma$ & 9.4\\
\hline
\multicolumn{4}{c}{optical lattice} \\
\hline
$\lambda_0^l$ & 1 $\mu$m &  $\lambda_0$ & $106$ nm\\
intensity & $1.6\cdot10^{16}$ W/cm$^2$  & $a_0$ & 0.11 \\
 ramp & 200 $\mu$m & $\tau_{inj}$ & 200 $\lambda_0^l/c$\\
\hline
\multicolumn{4}{c}{scattered light} \\
\hline
$\lambda_s^l$ & 5.6 nm & $\omega_s$ & 0.985 $\omega_0$\\
\hline \hline\noalign{\smallskip}
\end{tabular*}
\end{table}

\Cref{fig2} presents the time evolution of the scattered field amplitude on the right boundary. In the interval $100<t<
500 \lambda_0/c$, one can see an exponential growth of the signal, followed by saturation at the amplitude $\langle
a_s\rangle/a_0 \simeq 1.4\cdot10^{-3}$. The exponential growth rate is estimated as $\Gamma/\omega_0 = 1.9\cdot
10^{-3}$, the fit being plotted as a dotted line.  For 1 $\mu$m laser wavelength, this corresponds in the laboratory
frame to a gain length in intensity of only $L_\textrm{gain}^l = 84\,\mu$m.

The laser electric field changes its sign across the zero lines of ponderomotive potential wells. The beat between the
laser and the scattered wave is hence in phase opposition on the two sides of a well, resulting in an anti-symmetric
bunching structure. This is illustrated in \cref{fig3}, showing the completely trapped electron distribution at
saturation. A quasi-sinusoidal bunching shape can be noticed, in sharp contrast with the series of parallel
micro-bunches of conventional  free electron lasers.

These numerical findings can be completed by an analytical  analysis; while several theoretical approaches are possible,
we choose to put forward a hydrodynamic model, that yields relevant  estimates for the scaling laws within a simple
mathematical framework. We propose to model amplification analytically by describing the collective electron behaviour
$n(x,z,t)$ as an eigenmode of the light potential well. In a simplified approach we neglect the electrostatic potential
of electrons and the longitudinal temperature, and consider the transverse temperature not to exceed the trapping limit.
Electron motion follows hydrodynamic equations, coupled to the propagation equation for the light vector potential :
   \begin{subequations}\label{fluid_eqs}
  \begin{align}
  &\partial_t n_e + \nabla\cdot (n_e\,\mathbf{u}) = 0 \label{fluid_eqs1}
   \end{align}
   \begin{align}
 &  \partial_t u_x + \nabla_x P_{xx}/m_e n_e + \nabla_x U_p = 0\label{fluid_eqs2}
   \end{align}
   \begin{align}
 &  \partial_t u_z + \nabla_z U_p = 0\label{fluid_eqs22}
   \end{align}
   \begin{align}
&  \left((\partial_t-i\omega_0)^2 - c^2 \nabla^2\right) \bar{a} = - (4\pi e^2/m_e) n_e\, \bar{a}\label{em_field}
  \end{align}
   \end{subequations}
where $\mathbf{u}$ is the electron fluid velocity. Considering the slowly oscillating vector potential $\bar{a}$
defined earlier, the ponderomotive potential reads $U_p = m_e(c/2)^2|\bar{a}|^2$. Note that $\bar{a}$ in
\cref{fluid_eqs} is the total field, including the scattered and  laser fields.

The beam is trapped along $x$, so it follows the adiabatic equation of state with the electron pressure reading
\(P_{xx} = mv_\perp^2 n^3/3n_0^2\), where $v_\perp = c\delta\beta_{\perp0}$ is defined by the transverse velocity
spread. Considering equilibrium between lattice potential and electron pressure, we find the transverse density
distribution of the background state \[n_e = n_0\sqrt{1-\xi^2}\,,\] where $\xi=x/L$ is a coordinate normalized to
the beam width $L = v_\perp/\Omega$, and $n_0=n_e(0)$ is a maximum electron density.

\begin{figure}[!ht]\centering
 \includegraphics[width=0.45\textwidth]{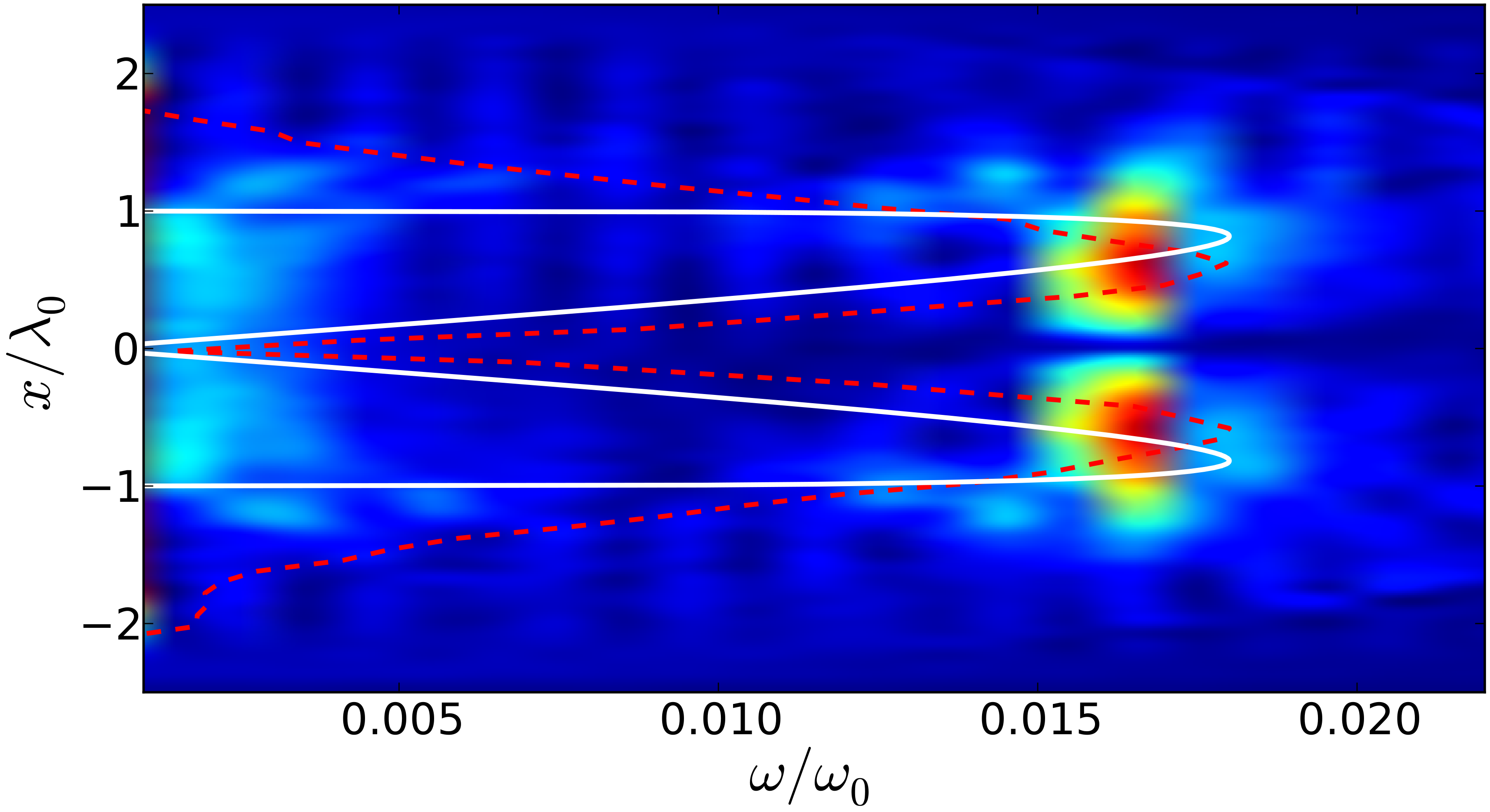}\\
 \caption{Spatial-spectral distribution of electron density perturbations. The predicted coupling shape $g(\xi)$ is
shown in white curve for $L = \lambda_0$ and the red dashed line is a slice of the spatial-spectral diagram along
$x$-axis at $\omega =\Omega$.}\label{fig4}
\end{figure}

In the linear approach, we consider the collective electron modes as a first-order perturbation of the beam position
$\Delta_x$, defined as $\xi = (x+\Delta_x)/L$. Without a signal wave there is no longitudinal force $\partial_z U_p$,
and from \cref{fluid_eqs} it follows: \[ \partial_t^2 \Delta_x = - \Omega^2\Delta_x \,,\] which describes the collective
beam oscillations with a frequency $\Omega$. 

The small-amplitude signal
wave $a_s$ is taken as a perturbation to the laser vector potential, inducing to first order a ponderomotive
potential $U_p^{(1)} = m_e(c/2)^2 \bar{a}_s \bar{a}_L^* + c.c.$. Assuming a plane scattered wave -- thus, neglecting
diffraction --,  we linearize  \cref{fluid_eqs} and turn to the Fourier domain, to obtain the dispersion equation:
\begin{align}\label{disp}
&\left(\omega^2 - \Omega^2 \right) \left( (\omega - \omega_0)^2 - (k_z-k_{0\parallel})^2c^2 \right) = \alpha
\omega_0^4\,.
\end{align}
This equation describes electron and electromagnetic modes with a coupling coefficient $\alpha = G (\omega_p a_0 k_z
c\,k_{0\perp} L/\omega_0^2)^2$, where $G$ is an overlap integral resulting from an inhomogeneous profile of the laser
field $\propto x$ and the beam density $\propto\sqrt{1-x^2/L^2}$. Thus, the model predicts that amplification is
localized mainly in the regions around the coupling maxima at $\xi=\pm\sqrt{2/3}$, with widths $\xi\simeq 0.5$. This
analytical prediction was checked numerically. \Cref{fig4} shows the spectra of electron density perturbations
(abscissa) as a function of the transverse position $x$, for an arbitrary fixed position $z$. The spectra are centered
at  the betatron frequency \(\Omega = 0.016\,\omega_0\). The transverse distribution consists of two symmetric
maxima, as predicted by the hydrodynamic model. The dashed line is obtained from the simulation, while the white curve
shows the transverse coupling function for a trapped beam half-width  $L = \lambda_0$. The numerical shape is slightly
broader, but the general structure of the emitting zones is well reproduced.

Solution to \cref{disp} for the Stokes mode is a complex frequency with a real part $\omega \simeq \Omega$ and a
positive imaginary part $\Gamma$, defining the amplification growth rate. Assuming a weak coupling condition $\alpha\ll
(\Omega/\omega_0)^3$ , the growth rate is:
\begin{equation}\label{grwth_rate_weak}
\Gamma/\omega_0 = 0.5 (\alpha\omega_0/\Omega)^{1/2}\,.
\end{equation}
Estimating $G \simeq 0.2$, and using parameters applied in numerical modelling (\cref{tab2}), we obtain a growth rate
$\Gamma/\omega_0 = 1.7\cdot 10^{-3}$. 

The exponential growth of the signal saturates when the electron beam gets trapped in the longitudinal potential
$|\bar{a}_s \bar{a}_L^*|$, resulting a full particle bunching $n_e \sim  n_0 \exp(-i\Omega t) +c.c.$. From
\cref{em_field}, the maximal signal amplitude can be deduced as :
\begin{equation}\label{saturat_lim}
\langle a_s\rangle /a_0 \simeq (k_{0\perp}L)\omega_p^2/(\Omega\omega_0)\,.
\end{equation}
For the chosen parameters, the signal saturation occurs at $\langle a_s\rangle/a_0 = 1.7\cdot10^{-3}$. These 
analytic estimates of growth rate and maximum of the scattered field are in good agreement with results of
numerical modelling.

In the laboratory frame, the growth rate reads $\Gamma^l/\omega_0^l = \Gamma/2\omega_0$ , and  can be expressed as a function of initial parameters:
\begin{equation}
\Gamma^l/\omega_0^l  = 0.1\, (K/\gamma)\,(\lambda_0/\sigma_\perp)\,\sqrt{a_0 \,J/J_A}
\end{equation}
where $K= k_{0\perp}L$ is a numerical factor defined by the filling of potential channel by electrons and $J_A = 17$ kA 
is the Alfven current. In normal trapping conditions, $K \simeq 1$.

These scaling laws provide estimates of the gain and saturation level from a Raman XFEL laser derived from
laser-wakefield acceleration. Let us consider parameters close to ones achievable in the experiments \cite{lundh} : an
electron bunch of 50 MeV, with a peak current of 14 kA, and 1 mm-mrad emittance. Twin undulator lasers are at a
wavelength of 800 nm, with an intensity of $3\cdot 10^{17}$ W/cm$^2$, corresponding to a normalized vector potential of
0.4. The output X-ray photons have an energy of 30 keV; the gain in intensity extrapolated from our numerical results is
15 cm$^{-1}$. Assuming a transverse size of 3 $\mu$m, an X-ray duration of 10 fs, results in an X-ray energy at 
$\mu$J level, or  $10^{8} - 10^{9}$ photons.

Note that photon emission results in an electron recoil. In the electron rest frame, the recoil can be observed
numerically as a collective backwards motion along $z$-axis, which starts near the saturation stage. Since emission is a
quantum process, in the case of high photon energies the recoil  has a discrete nature, and a quantum description may
become appropriate. A significance of quantum effects is defined by a quantum-recoil parameter \cite{bonifacio}, as a
ratio of the recoil induced detuning and  the gain bandwidth. It reads in our model $q_\lambda' = (\gamma_e
\hbar\omega_0)/(m_e c^2\Gamma/\omega_0 )$. Parameters in the present examples are restricted to $q_\lambda' <1$, where a
classical description is valid.

In summary, we have shown that a relativistic electron bunch, injected into and guided by a high intensity optical
lattice, triggers a forward Raman instability resulting in a  rapid amplification of a coherent X-ray beam. If coupled
to laser-wakefield acceleration, this process promises to yield  hard X-ray laser beams. Raman
processes often exhibit an intrinsic robustness with respect to random variations of the driving parameters
\cite{kruer}, such as the laser intensity, which may be a key to an experimental demonstration. The Raman X-ray free
electron laser should therefore be further studied as an  opportunity to supply ultra-compact coherent X-ray sources, up
to the hard X-ray range.


\end{document}